\documentclass{n}
\usepackage{amsmath,amssymb,graphicx,hyperref}    
\title{\large Differential Neutrino Condensation onto Cosmic Structure}

\author{
  Hao-Ran Yu$^{1,2,3}$,
  J.D. Emberson$^{3,4}$,
  Derek Inman$^{3,5}$,
  Tong-Jie Zhang$^{1\ast,6}$,
  Ue-Li Pen$^{3,7,8,9}$,
  Joachim Harnois-D\'{e}raps$^{10,11}$,
  Shuo Yuan$^{12}$,
  Huan-Yu Teng$^{1}$
  Hong-Ming Zhu$^{13}$,
  Xuelei Chen$^{13}$,
  Zhi-Zhong Xing$^{14,15}$,
  Yunfei Du$^{16,17}$,
  Lilun Zhang$^{18}$,
  Yutong Lu$^{16,17}$
  \& XiangKe Liao$^{16}$\\
  {\footnotesize $\ast$ tjzhang@bnu.edu.cn}
  }

\begin{document}

\maketitle

\begin{affiliations}
\footnotesize{
\item Department of Astronomy, Beijing Normal University, Beijing 100875, China
\item Kavli Institute for Astronomy \& Astrophysics, Peking University, Beijing 100871, China
\item Canadian Institute for Theoretical Astrophysics, University of Toronto, M5S 3H8, Ontario, Canada
\item Department of Astronomy \& Astrophysics, University of Toronto, Toronto, Ontario M5S 3H4, Canada
\item Department of Physics, University of Toronto, Toronto, Ontario M5S 1A7, Canada
\item Shandong Provincial Key Laboratory of Biophysics, School of Physics and Electric Information, Dezhou University, Dezhou 253023, China
\item Dunlap Institute for Astronomy and Astrophysics, University of Toronto, Toronto, ON M5S 3H4, Canada
\item Canadian Institute for Advanced Research, Program in Cosmology and Gravitation
\item Perimeter Institute for Theoretical Physics, Waterloo, ON, N2L 2Y5, Canada
\item Department of Physics \& Astronomy, University of British Columbia, Vancouver, B.C. V6T 1Z1, Canada
\item Scottish University Physics Alliance, Institute for Astronomy, University of Edinburgh, EH9 3HJ, Scotland, UK
\item Department of Astronomy, Peking University, Beijing 100871, China
\item Key Laboratory for Computational Astrophysics, National Astronomical Observatories, Chinese Academy of Sciences, Beijing 100012, China
\item School of Physical Sciences, University of Chinese Academy of Sciences, Beijing 100049, China
\item Institute of High Energy Physics, Chinese Academy of Sciences, Beijing 100049, China
\item School of Computer, National University of Defense Technology, Changsha 410073, China
\item National Supercomputer Center in Guangzhou, Sun Yat-Sen University
\item Institute of Ocean Science and Technology, National University of Defense Technology}
\end{affiliations}

\begin{abstract}
Astrophysical techniques have pioneered the discovery of neutrino
mass properties. Current cosmological observations give an upper
bound on neutrino masses by attempting to disentangle the small
neutrino contribution from the sum of all matter using precise
theoretical models. We discover the differential neutrino
condensation effect in our TianNu $N$-body simulation.
Neutrino masses can be inferred using this effect by
comparing galaxy properties in regions of the universe with
different neutrino relative abundance (i.e. the local neutrino
to cold dark matter density ratio). In ``neutrino-rich'' regions,
more neutrinos can be captured by massive halos compared to
``neutrino-poor'' regions. This effect differentially skews
the halo mass function and opens up the path to
independent neutrino mass measurements in current or future
galaxy surveys.
\end{abstract}

Neutrinos are elusive elementary particles whose fundamental
properties are incredibly difficult to measure.
40 years after their first direct detection
\cite{1956Sci...124..103C,PhysRevLett.9.36},
flavour oscillation experiments \cite{PhysRevLett.81.1562,
2002PhRvL..89a1301A,PhysRevLett.87.071301}
confirmed that at least two neutrino types are massive and
placed a lower bound on the sum of their mass:
$M_\nu \equiv \sum_{i=1}^3 m_{\nu_i} \gtrsim$ 0.05 eV
\cite{2014ChPhC..38i0001O}.
This discovery has a profound impact on our
understanding of the early Universe,
where neutrinos are produced in great numbers.
First in a relativistic state, they contribute to
the radiation energy density,
thereby modulating the matter-to-radiation ratio in
a way that depends on their mass. This leaves an
imprint on the Cosmic Microwave Background (CMB) 
that can be searched for, providing a current  upper bound of
$M_\nu \lesssim$ 0.23 eV \cite{2015arXiv150201589P}.
As the Universe expands,  the neutrinos lose momentum
and eventually become non-relativistic, at which
point they contribute instead to the matter energy density.
Although small, this neutrino component has consequences 
on the clustering properties of matter, which can be
measured in observations of the Large Scale Structure (LSS).
Despite having a current temperature of roughly 2 K
today, these relic neutrinos maintain a velocity dispersion
that is significantly higher than both baryons and
cold dark matter (CDM). This thermal motion (free
streaming) reduces their susceptibility to
gravitational capture, leading to suppressions in
the matter power spectrum. Upcoming cosmological
missions that probe the large-scale structure of the
Universe, such as Euclid \cite{2011arXiv1110.3193L}, LSST
\cite{2009arXiv0912.0201L}, eBOSS \cite{2015arXiv150804473D}, and DESI
\cite{2015AAS...22533605E}, will complement CMB measurements and
are expected to improve this upper bound in the near future
\cite{2015APh....63...66A}. Unfortunately, it is extremely difficult to
disentangle this suppression from differences in modelling of the
non-linear gravitational growth and galaxy/halo bias,
the uncertain impact of baryonic processes and potentially
alternate models of gravity \cite{2016MNRAS.tmp..468M}.

While there have been some attempts to study the interplay between
neutrinos and LSS analytically \cite{2014PhRvD..90h3518L},
the current understanding of their non-linear dynamics heavily relies 
on cosmological $N$-body simulations that coevolve both CDM and
neutrino particles (e.g. \cite{2015JCAP...07..043C}).
These allow to investigations of, for example,
the effects of massive neutrinos on the halo mass function and
on the galaxy bias, or to study the neutrino density profile
\cite{2010JCAP...09..014B,2014JCAP...02..049C,2013JCAP...03..019V}.
The precision of these subtle measurements is limited both by
the simulation volumes and Poisson noise, and is challenging
even for modern high performance computing centres.

Taking advantage of a development phase of the Tianhe-2
supercomputer, we have completed the world's largest cosmological
$N$-body simulation, named ``TianNu'', which coevolved
just under 3 trillion CDM and neutrino particles. From this
calculation we find that regions with same CDM density
can have substantially different mean neutrino densities
due to
the free streaming of neutrinos. Consequentially,
collapsed dark matter halos that live in ``neutrino-rich''
sectors of the Universe have deeper gravitational potential
wells to trap even more CDM and baryons, compared to
otherwise identical halos from ``neutrino-poor'' regions.
This is a novel form of feedback on the cosmic structure
whose strength depends on $M_{\nu}$, which we refer to as
{\it differential neutrino condensation}. It results in
spatial variations of halo properties that depend on the
{\it difference} between the neutrinos and CDM rather than
on their {\em sum}, and as such provides a clean probe of
$M_{\nu}$ than conventional LSS probes.

\paragraph*{(Simulation.)}
\begin{figure} \centering
  \includegraphics[width=1.0\linewidth]{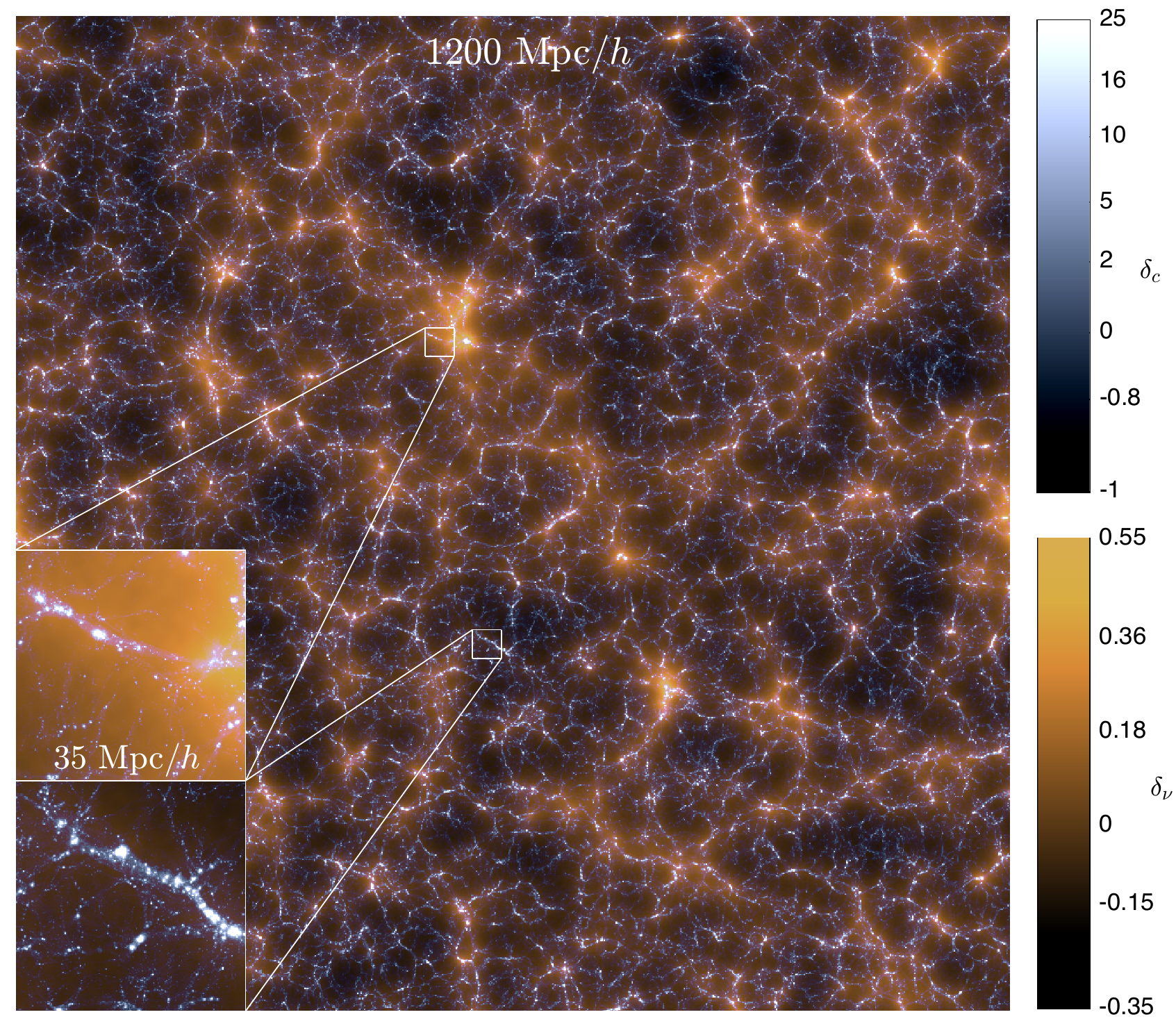}
  \caption{\textbf{TianNu simulation.}
    Two-dimensional visualisation of CDM and neutrino structures
    in TianNu. CDM is represented in blue-white and neutrinos in orange.
    The two subpanels focus on regions with similar CDM structure
    but different neutrino-to-CDM density ratios. The upper panel shows a
    neutrino-rich region with $(\delta_c,\delta_\nu)=(0.081,0.19)$
    while the lower panel shows a neutrino-poor region for which
    $(\delta_c,\delta_\nu)=(0.44,-0.047)$. Density contrasts are
    averaged over the volume of each subpanel, $35^2\times 8.3\ ({\rm
    Mpc}/h)^3$.  The difference in neutrino condensation seen in these
    two panels leads to systematically different halo properties
    between the two regions.}
  \label{fig.1}
\end{figure}
TianNu\footnote{See
\url{http://www.cita.utoronto.ca/~haoran/thnu/movie.html}
for a visualisation of the two-component evolution in TianNu.}
was carried out with a modified version of the public cosmological
$N$-body code {\sc cubep$^3$m} \cite{2013MNRAS.436..540H} that
includes CDM and neutrino coevolution \cite{2015PhRvD..92b3502I}.
The neutrino sector in TianNu models the minimal
``normal hierarchy'' of the neutrinos,
in which 2.046 massless neutrino species are included in the
background cosmology at the level of the initial transfer function
\cite{2011JCAP...07..034B}, and one $m_\nu = 0.05$ eV neutrino
species is explicitly traced using $N$-body
particles. The background cosmological environment is otherwise
described by the densities of CDM and baryons, Hubble's parameter,
the initial tilt and fluctuations of the power spectrum, which
are parameterised with  [$\Omega_c, \Omega_b, h, n_s, \sigma_8$] 
= [0.27, 0.05, 0.67, 0.96, 0.83]. We finally impose flatness by
setting $\Omega_\Lambda=1-\Omega_m$, where the total matter
density is $\Omega_m=\Omega_b+\Omega_c+\Omega_\nu$. $6912^3$ CDM
and $13824^3$ neutrino $N$-body particles are coevolved in a
cosmological volume of $L=1.2$ ${\rm Gpc}/h$ on the side.
which is comparable to a shallower,
lower redshift galaxy survey, in which neutrino effects are
relatively prominent. For 
comparison and calibration purposes, we ran a neutrino-free
simulation (equivalent to a TianNu simulation in which all
$m_\nu = 0$ eV) with same initial condition and $\Omega_m$ fixed,
named ``TianZero''.

Fig.1 shows a two-dimensional projection of a thin slice of
thickness $8.3$ ${\rm Mpc}/h$ extracted from the TianNu
simulation at $z = 0.01$. The blue-white colour scale shows
the CDM component $\delta_c$, while orange highlights the neutrinos,
$\delta_{\nu}$. Here $\delta_i$ are the density contrasts
obtained from the density fields $\rho$ with
$\delta_i \equiv \rho_i/\bar{\rho_i}-1$. $\delta_c$ exhibits
the ``cosmic web'' structure, mostly arranged into filaments
and collapsed halos. On scales $k^{-1}>100$ ${\rm Mpc}/h$, which
is much larger than neutrino's free streaming scale, $\delta_c$
and $\delta_\nu$ are coupled via their mutual gravitational
potential and have similar structures.
However, due to the free streaming, on smaller scales $k^{-1}<20$
${\rm Mpc}/h$, neutrinos are instead distributed as diffuse
``condensation clouds'' centered on the largest CDM
structures and so $\delta_\nu$ is less correlated
with $\delta_c$. We find that, this environmental difference
between $\delta_c$ and $\delta_\nu$ on the scale
$k^{-1}\simeq 10$ ${\rm Mpc}/h$, causes systematic modulations
on CDM halo (of scale $k^{-1}\simeq 2$ ${\rm Mpc}/h$) properties.
The two subpanels in Fig.1 provide a visual
representation of the large variations in the local difference
of $\delta_c$ and $\delta_\nu$ in regions with otherwise similar LSS. 

\paragraph*{(Halo masses.)}

In order to demonstrate how the properties of galaxies are affected
by differences in their neutrino environment, we first search for
collapsed structures in  both TianNu and TianZero. We identify
spherical CDM halos, which are the regions that host galaxies.
For each halo, we record the total mass $m_h$ and position $\vec{x}$,
rejecting under-resolved halos with mass $m_h<3.5\times
10^{11}~M_\odot$. We do not include
neutrinos in the evaluation of $m$ since the observable quantity
that we measure -- the luminosity of its constituent galaxies -- 
correlates only with the CDM mass. Since both simulations have
the exact same CDM initial conditions, systematic differences in
their halo properties are caused by the neutrino mass. The first
step in this comparison consists in matching halos in
TianZero with their counterpart in TianNu, representing the
evolution of the same physical object in different cosmologies. 
Then we sort all halos in descending order according to their
masses, and assign a rank, $r$, to each halo, defined as its index
in the sorted list. This mass rank serves as a proxy for
observational properties of galaxies (e.g. by abundance matching
\cite{2011ApJ...742...16T,2013arXiv1310.3740K,2015MNRAS.447.3693K},
halo occupation distribution
\cite{2002ApJ...575..587B, 2005ApJ...633..791Z}).
The ranks $r_\nu$ and $r_0$ obtained from TianNu and TianZero
are in general different, which means that the presence of
massive neutrinos affects the final mass of collapsed
structures in a way that depends not only on the halo mass -- the
rank would be unchanged --  but also on their environment. 

We next estimate the forementioned local values of $\delta_c$ and
$\delta_\nu$ centered on the position of each halo, by a
reconstruction technique \cite{2015PhRvD..92b3502I} that estimates
the CDM and neutrino density fields directly from the halo
catalogues. The resulting quantities, $\hat\delta_c$ and
$\hat\delta_\nu$, on scale $k^{-1}\simeq 10$ ${\rm Mpc}/h$, describe
the two environment variables that are responsible for the changes
of halo ranks between TianNu and TianZero. These can be extracted
from galaxy redshift surveys, thereby leading to a measurable
neutrino effect. Neutrino mass has effect primarily
on halo masses rather than halo positions, so halo catalogue
from either TianNu and TianZero gives same estimation of
$\hat\delta_c$ and $\hat\delta_\nu$. We examined that the
results are insensitive to the reconstruction accuracy.

\begin{figure} \centering
  \includegraphics[width=0.6\linewidth]{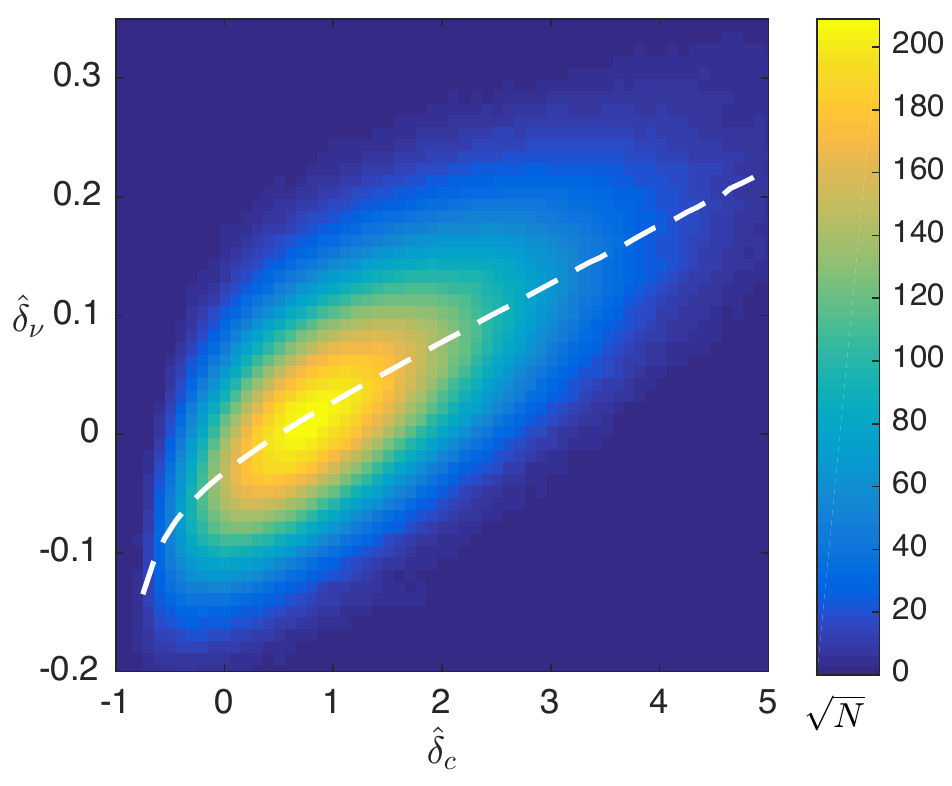}
  \caption{\textbf{Distribution of dark matter halos as a
  function of environmental variables $\hat\delta_c$ and
  $\hat\delta_\nu$.}
  This figure exhibits the correlation between the two
  quantities, and the considerable amount of spread.
  The dashed line shows the mean value of $\hat\delta_\nu$
  in each $\hat\delta_c$ bin, which serves as a baseline to
  define the ``neutrino excess'' $\epsilon_\nu$.}
  \label{fig.2}
\end{figure}
In Fig.2 we plot the $\hat\delta_c$ and $\hat\delta_\nu$'s
distribution of halos. The correlation between the two
quantities (with a Pearson's $r$ $\rho_{c,\nu}=0.663$)
shows their coupling above neutrino's free streaming
scale. Namely, regions of high (low) $\hat\delta_c$ have
{\it generally} high (low) $\hat\delta_\nu$. This correlation
is undesired, as any observable dependence on $\hat\delta_\nu$
is entangled with $\hat\delta_c$. However,
the considerable amount of spread in Fig.2 indicates that halos residing
in identical $\hat\delta_c$ regions {\it need not} have same
$\hat\delta_\nu$ environments. In order to disentangle the {\it net}
neutrino's contribution from CDM's contribution, we
subtract the mean correlation and consider only the scatter
of $\hat\delta_\nu$ at a given $\hat\delta_c$, i.e., for a given $\hat\delta_c$, we
measure the expectation values of $\hat\delta_\nu$, $\langle
\hat{\delta_{\nu}} \rangle$, indicated with the dashed
line in the figure. From this quantity we define the
``neutrino excess'', $\epsilon_\nu \equiv \hat{\delta_{\nu}}
- \langle \hat{\delta}_{\nu} \rangle$, which is uncorrelated
with $\delta_c$. A halo with
$\epsilon_\nu>0$ is located in a region with relatively
high neutrino density; and we refer it to
``neutrino-rich'' region. On the contrary, $\epsilon_\nu<0$
refers to ``neutrino-poor'' regions. This is a quantitative
measure of {\it difference} of the CDM and neutrino abundance
caused by differential neutrino condensation.

The dominant effect of this differential condensation is to
alter the rank ordering of halos. To illustrate this,
we measure the relative variation of rank between the TianNu
and TianZero simulations, $\Delta r/r\equiv (r_\nu-r_0)/r_0$,
measured from each halo pair, and organise the results as a
function of neutrino excess and rank. This is shown in Fig.3,
where the colour scale indicates the expectation value of 
the relative change in rank as a function of rank and
$\epsilon_\nu$. We rescale each pixel by a weight function
$w=\sqrt{N}$, where $N$ is the number of halos in each pixel.
The top axis provides a conversion from rank to halo mass.
We can see that
the neutrino-rich (upper part) regions are bluer, meaning
that the mass of halos in TianNu tend to be systematically
{\it increased} (the rank decreases) by the presence of 
high neutrino condensation compared to the TianZero baseline.
Similarly, while the neutrino poor (bottom part) regions
are redder, corresponding to a systematic {\it decrease}
of the halo mass in neutrino-poor regions. This is a
manifestation of ``differential neutrino condensation'', and
its impact on cosmic structure is the strongest for
cluster-scale halos ($m_h\gtrsim 10^{13}M_\odot$). It is a
characteristic signature of the gravitational feedback of
modulations in the neutrino density onto the growth of
CDM halos. 
\begin{figure} \centering
  \includegraphics[width=0.6\linewidth]{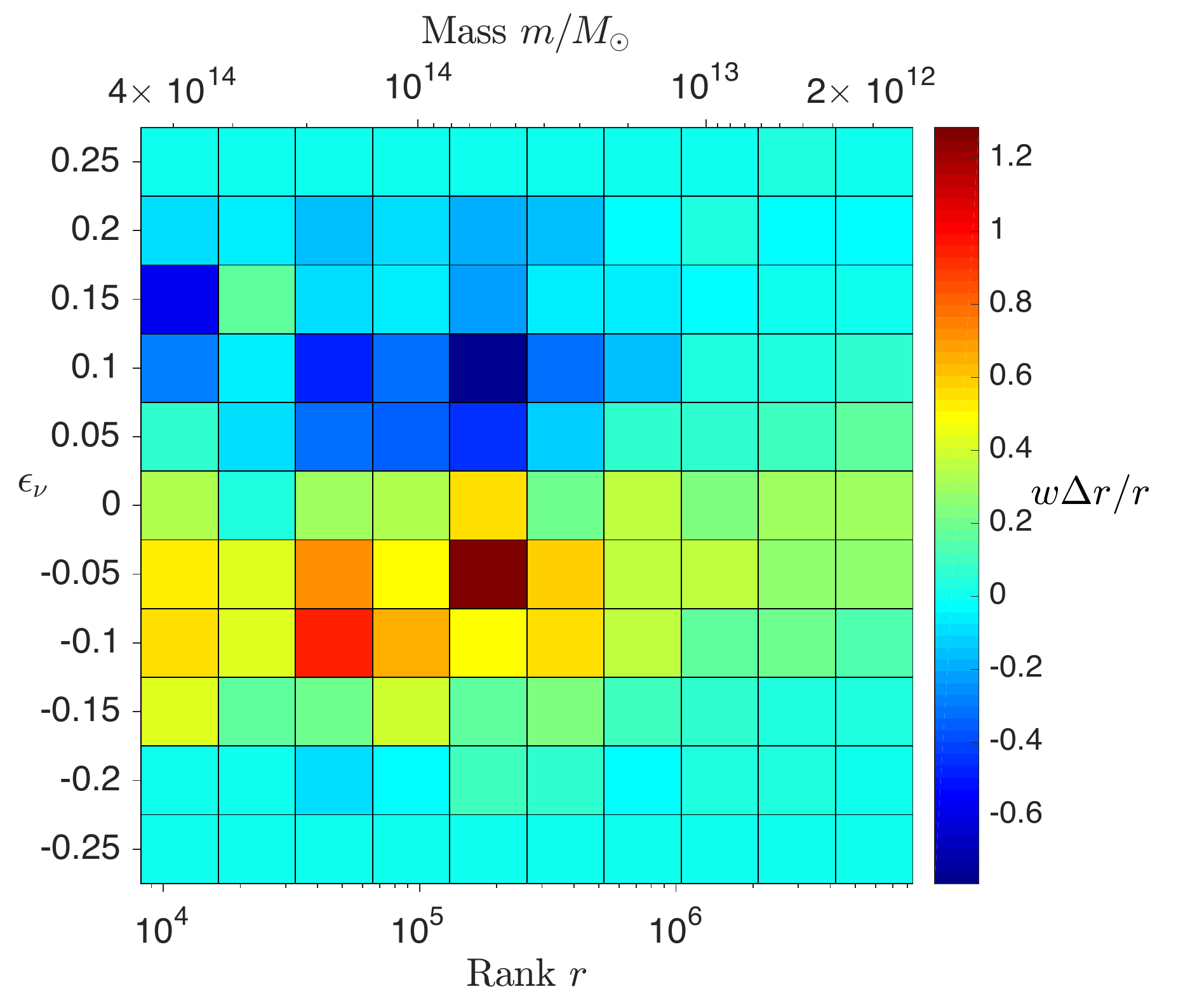}
  \caption{\textbf{Relative variation in halo mass and rank
  caused by differential neutrino condensation.}
  The colour scale shows a weighted histogram of the 
  relative rank variations observed between TianNu and
  TianZero, $\Delta r/r\equiv (r_\nu-r_0)/r_0$, organized
  in bins of halo rank (or mass) and neutrino excess
  $\epsilon_\nu$. More massive halos (lower rank) that
  live in neutrino-rich regions condense enough neutrinos
  to lower their ranks and increase their mass. Conversely,
  in neutrino-poor regions, equally large halos lose mass
  due to neutrino free streaming, increasing their rank. 
  This effect becomes weaker for less massive halos
  since these are less efficient in condensing neutrinos.}
  \label{fig.3}
\end{figure}

The global trends in rank variations can be expressed by a
linear function of $r$ and $\epsilon_\nu$. The rank is
first held fixed, and the variations in $\Delta r/r$
with changing $\epsilon_\nu$ (e.g. columns in Fig. 3)
are fit with a straight line. The slope of this line,
$\partial(\Delta r/r)/\partial\hat{\delta}_\nu$,
is shown in Fig.4 for different rank bins. We repeat this
measurement from 1000 bootstrap resamplings of the halo
catalogues in order to estimate the error bars. These
points are all negative, indicating that for a halo of
rank $r$, a net increase of neutrino density
$\partial\hat{\delta}_\nu$ lowers the rank by
$\partial(\Delta r/r)$. The effect asymptotes to zero as
the halo mass decreases, reflecting the results seen in
the rightmost region of Fig.3. On the high-mass end, the
measurements are well described by $\partial(\Delta r/r)/\partial
\hat{\delta}_\nu=\alpha\log_{10}(r/r_{\rm pivot})+\beta$,
with $\{\alpha,\beta\}=\{0.00254\pm 0.00073,-0.00194\pm 0.00017\}$, 
shown as the red dashed line in Fig.4. The value $r_{\rm pivot}
=8.1\times 10^5$ is chosen such that $\alpha$ and $\beta$ are
uncorrelated.
\begin{figure} \centering
  \includegraphics[width=0.6\linewidth]{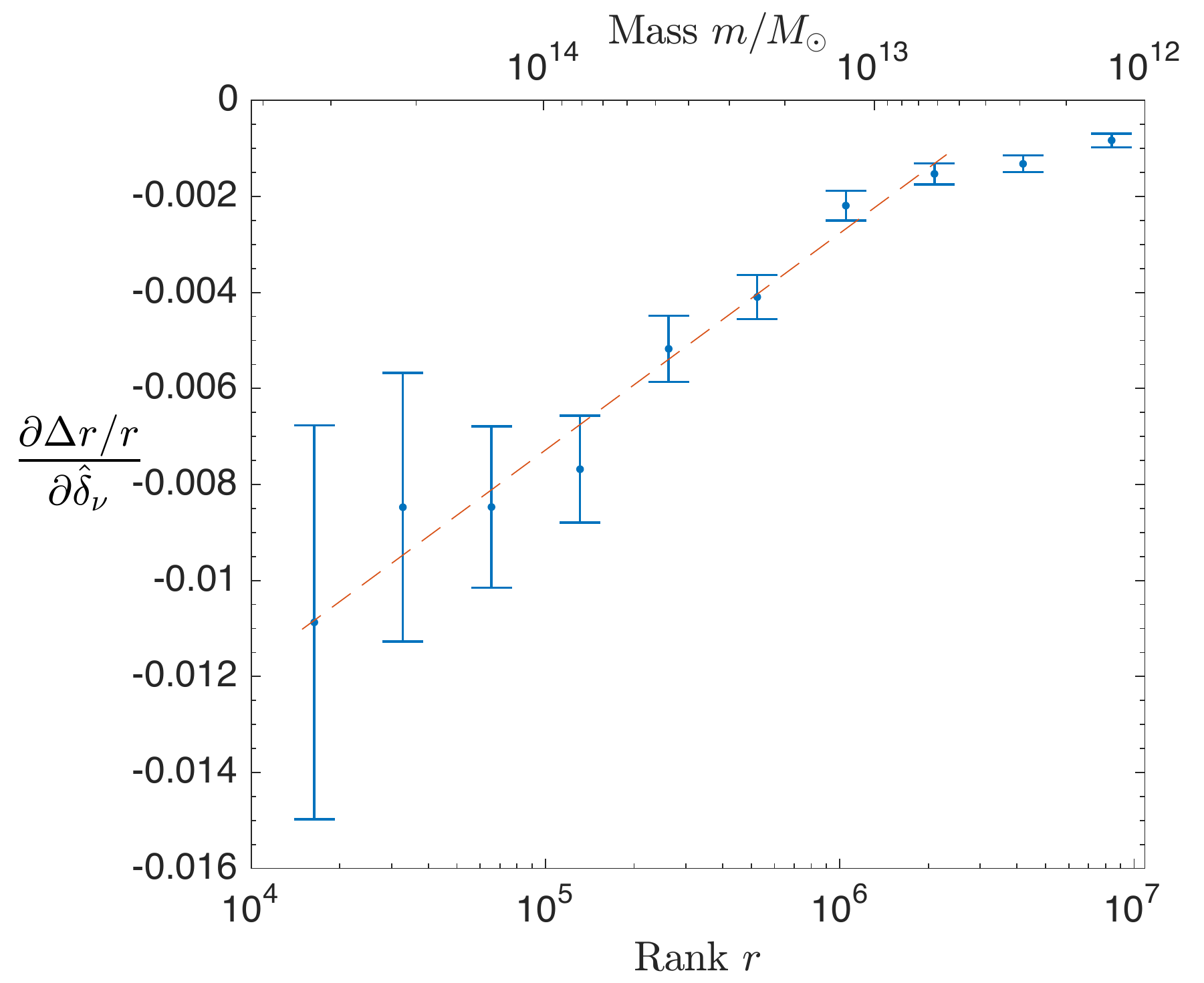}
  \caption{\textbf{Differential modulation in halo rank caused by
  neutrino condensation.}
  The measurements are shown as a function of halo mass (and rank),
  and the error bars are from bootstrap resampling.
  All points are negative with strong signal to noise,
  indicating that differential neutrino condensation has an
  impact on all halos. The effect is stronger for high
  mass halos (low rank) since they can trigger neutrino
  condensation in a more effective way, provided they are in
  a region of high $\epsilon_{\nu}$. The red dashed line
  represents the best linear fit to the measurement, for mass
  $m>5\times 10^{12} M_{\odot}$.}
  \label{fig.4}
\end{figure}

\paragraph*{(Discussion and conclusion.)}

The TianNu and TianZero simulation demonstrate a controlled
numerical experiment to show the differential neutrino
condensation effect. In order to isolate neutrino effects
while cancelling CDM and baryon contributions to halo
properties, we perform three differences. First,
we use neutrino excess $\epsilon_\nu$, a quantity uncorrelated
with CDM density, to distinguish regions in our simulation that
are neutrino-rich and neutrino-poor, according to the
local (on scale $k^{-1}\simeq 10$ ${\rm Mpc}/h$)
{\it difference} between $\delta_c$ and $\delta_\nu$.
Second, we search for halos' mass/rank differences between
TianNu ($M_\nu=0.05$ eV) and TianZero ($M_\nu=0$ eV), and
find their dependence on the neutrino excess $\epsilon_\nu$.
This dependence, written in $\partial(\Delta r/r)/\partial
\hat{\delta}_\nu$ relies on the {\it net} change of neutrinos density,
and is cleanly separated from $\delta_c$-$\delta_\nu$'s
correlation and potential baryonic effects.
Finally, we investigate how this dependence changes with halo
mass and find that the mass/rank of heavier halos are more
significantly affected by a net change of neutrino density.
The next step is to determine how this can be robustly
measured in astronomical observations.  The analysis in this report
focussed on plausibly observable quantities, specifically neutrino and
dark matter densities $\hat\delta_c$ and $\hat\delta_\nu$
reconstructed from halo catalogs fields, and rank order $r$ as proxy
for luminosity or number count observables.  The clean separation of
cosmological regions into neutrino poor and rich needs to be carefully
controlled to cancel generic dark matter contributions or 
potential super-sample variance \cite{2014PhRvD..89h3519L} to halo
properties.

\begin{methods}
Methods and numerical details are available in the
supplemental material.
\end{methods}



\section*{References}
\bibliographystyle{nm}
\bibliography{mybib}

\begin{addendum}
 \item The TianNu and TianZero simulations were performed
on Tianhe-2 supercomputer at the National Super
Computing Centre in Guangzhou. The analysis were
performed on the GPC and BGQ supercomputer at the
SciNet HPC Consortium. 
This work was supported by the National Science
Foundation of China
(Grants No. 11573006, 11528306, 10473002, 11135009),
the Ministry of Science
and Technology National Basic Science program
(project 973) under grant No. 2012CB821804,
the Fundamental Research Funds for
the Central Universities.
JHD acknowledge support from the European Commission under
a Marie-Sk{\l}odwoska-Curie European Fellowship (EU project
656869) and from the NSERC of Canada.
We thank Prof. Yifang Wang of IHEP for his great initial
support for our project, and Prof. Xue-Feng Yuan for
his kindly great support in Tianhe-2 supercomputing center.
HRY acknowledges General Financial Grant from the China Postdoctoal Science Foundation No.2015M570884.
JDE and DI acknowledge the support of the NSERC.

\item[Additional information]

\item[Competing financial interests] The authors declare that they have no
competing financial interests.

\end{addendum}

\clearpage

\section*{Methods}
This section contains details of the simulation and data analysis.
\paragraph{Code details.}
{\sc cubep$^3$m} is a publicly-available high performance cosmological
$N$-body code \cite{2013MNRAS.436..540H}.
It is written in {\tt FORTRAN} and created for
performance and scaling on highly parallelised supercomputers.
It is hybrid parallelised using MPI (node-level) and OpenMP (core-level)
to allow for optimal usage of multi-core nodes with further
parallelisation allowable via Nested OpenMP. Its force calculation
is done by a two-level particle mesh (PM) algorithm plus an adjustable
particle-particle pairwise force. The long range gravitational force
is computed on the node-level by solving Poisson's equation in Fourier
space using a 3D FFT with pencil decomposition \footnote{
D. Pekurovsky, P3DFFT: a framework for parallel computations of Fourier transforms in three dimensions, SIAM Journal on Scientific Computing 2012, Vol. 34, No. 4, pp. C192-C209.}
 for maximum geometric
flexibility. The short range force and particle-particle force are
computed on the core-level. Particle positions and velocities are
updated at each time step via a Runge-Kutta integration scheme.
Full details are provided in \cite{2013MNRAS.436..540H}.

\paragraph{Simulation details.}
The TianNu simulation uses $24^3=13824$ computing nodes ($331776$
cores) cubically distributed. The simulation scale is set to $L=1.2$
${\rm Gpc}/h$, containing $27648^3$ fine mesh cells and $6912^3$ coarse
mesh cells evenly distributed on the volume. Particle pairwise forces
are computed within fine cells with a softening length 0.3 times the
length of the fine cell ($\simeq 13 {\rm kpc}/h$). The total numbers
of CDM and neutrino particles are $6912^3$ and $13824^3$, and their
mass resolutions are $7\times10^8$ $M_\odot$ and $3\times10^5$
$M_\odot$ respectively. Our choice of neutrino mass models the
minimal ``normal hierarchy'' with two light species included in the
background cosmology by using {\sc class} \cite{2011JCAP...07..034B}
transfer function, and one heavy $m_\nu=0.05$ eV species whose
dynamics are explicitly traced using particles. The first step
in both TianNu and TianZero simulations involves the evolution of
$6912^3$ CDM particles from redshift $z=100$ to $5$. $13824^3$
Neutrino particles are subsequently added into the mixture at $z=5$
of TianNu, with $\Omega_m$ fixed\footnote{
Keeping $\Omega_m$ fixed guarantees the conservation of the total matter
power at scales much larger than the neutrino free streaming
scale, and the two simulations differ only on medium and small
scales, among which the analysis of environmental $\epsilon_\nu$
(medium scale)'s contribution to halos (small scale) is clean.
}
, and the two components evolve
under their mutual gravity to $z=0$. TianZero evolve only $6912^3$
CDM particles, equivalent to another TianNu simulation but setting
all $m_\nu=0$ eV. In total, TianNu and TianZero took 52 and 32
hours (17 and 11 million CPU hours) computation time on Tianhe-2.

\begin{figure} \centering
  \includegraphics[width=0.6\linewidth]{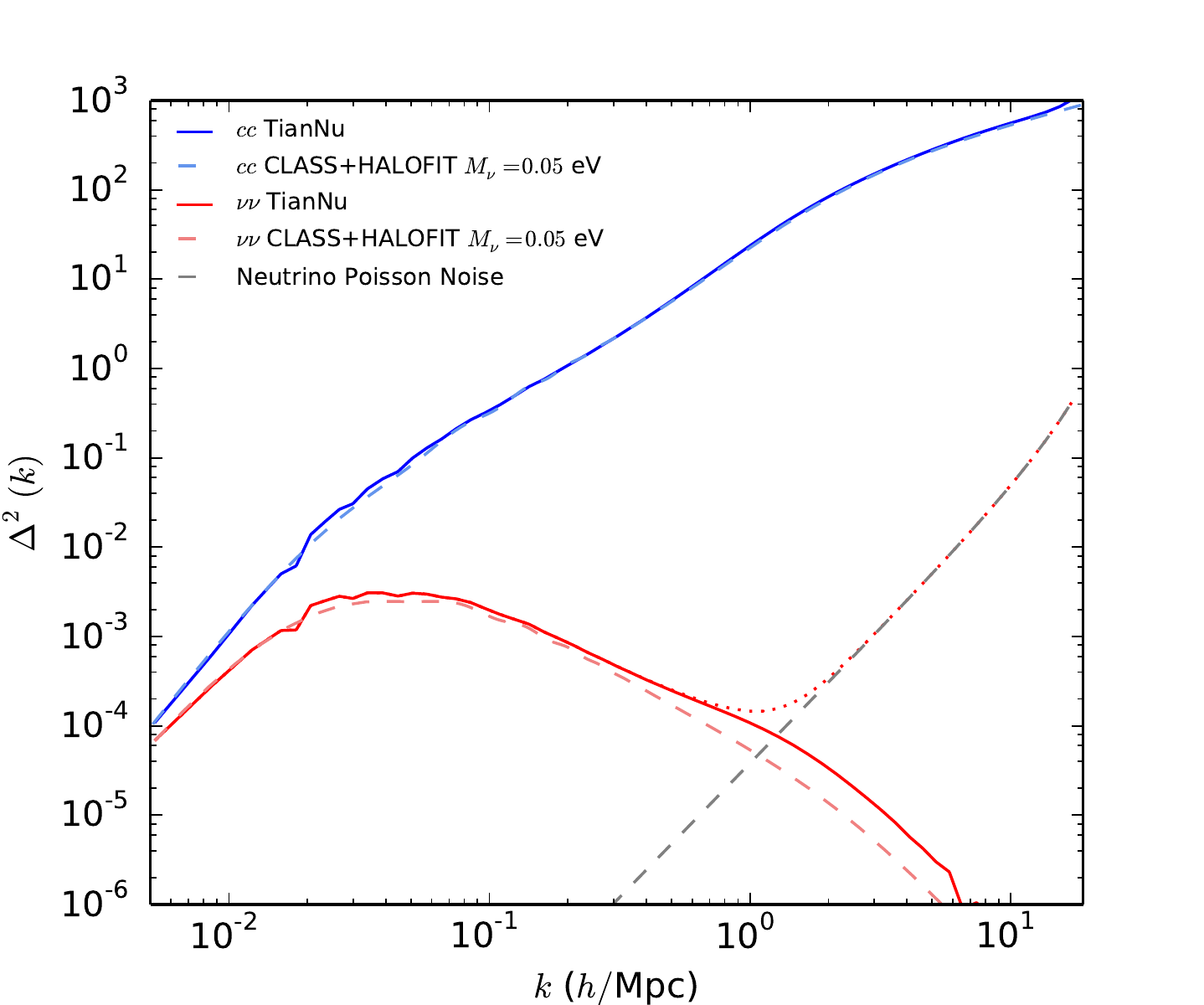}
  \caption{Dimensionless power spectrum $\Delta^2$ of $\delta_c$ and
  $\delta_\nu$ from the TianNu simulation and their nonlinear
  predictions. Poisson noise of neutrino particles shown in grey
  dashed line.}
  \label{fig.6}
\end{figure}
We output 21 checkpoints containing all CDM and neutrino positions
and velocities.  For our checkpoints, we store particles ordered by
their coarse cell location. We then record the number of particles
per coarse cell as a 1-byte integer, their fractional location across
the cell as a 1-byte integer (e.g. for a coordinate that is 24.4883
$\simeq$ 24 + 125/256 we store ``125'') per dimension. For velocities
we record one 2-byte integer per dimension indicating the fraction
of the maximum particle velocity with a resolution of $2^{-16}$.
In total, CDM requires 10 bytes per particle and neutrinos 9.125
bytes per particle as opposed to the traditional 24 bytes per
particle each (e.g. 6 coordinates each as a 4-byte float).
Analysis and results in this paper are based on the checkpoint
at redshift $z=0.01$. In Fig.5 we show the dimensionless power
spectrum $\Delta^2\equiv k^3 P(k)/2\pi^2$ of $\delta_c$ and
$\delta_\nu$ in TianNu at $z=0.01$, as well as their theoretical
predictions
\footnote{Given by $P_{ii}(k)=(T_i/T_m)^2 P_{mm}^{\rm HF}(k)$,
  where $i=c,\nu$,
  and $P_{mm}^{\rm HF}(k)$ is the power spectrum from {\sc halofit},
  and $T_i$ are the {\sc class} transfer functions.
  }, where $P(k)$ is the power
spectrum. One can see that the peak maximum at $k\simeq 0.04$ $h/{\rm
Mpc}$ in the neutrino power spectra -- corresponding to the neutrino free
streaming scale -- is well recovered both by the nonlinear prediction
(red dashed line) and in the TianNu simulation (red solid line,
with predicted Poisson noise subtracted). Another two scales of
interest are $k\simeq 0.01$ $h/{\rm Mpc}$ where $\delta_\nu$
still traces $\delta_c$ and $k\simeq 0.5$ $h/{\rm Mpc}$ where
the formations of halos take place. Our choice of simulation scale $L$
and enough neutrino particles prevent these three scales of interest
from being contaminated by neutrino Poisson noise and being affected
by the periodic boundary conditions of the simulation. Although
baryonic physics is non-negligible on and below halo scales, 
neutrinos in a halo come from $\sim 100$ ${\rm Mpc}/h$, while
CDM and baryons all come from $\sim 1$ ${\rm Mpc}/h$. Interaction
of CDM and baryons should not affect neutrinos.
Thus, the relative neutrino-rich and -poor properties is conserved
and $\partial(\Delta r/r)/\partial\hat{\delta}_\nu$ should be
unaffected. Additionally, the overall nature of our ``differential''
analysis cancels out the CDM and baryon effects, which are
uncorrelated with neutrino excess, $\epsilon_\nu$.

\paragraph{Halo finding and matching.}
The halo finding procedure uses a spherical overdensity approach.
It searches for local maxima in a nearest-grid-point CDM density
field with a threshold $\delta_c>100$. It then uses a parabolic
interpolation to search for the precise maxima of the particles.
These maxima are then inspected, largest $\delta_c$ first, and halo
mass is accumulated in spherical shells surrounding the maxima until
the mean density drops below $178$ times the mean density. After
selection as a halo, the mass is removed to avoid double counting in
the case of nearby maxima. For each halo, we record the total
mass of the CDM particles contained within its radius
\footnote{The
  radius of each halo is taken as $r_{200}$, the radius within which
  the density of the halo is 200 times the critical density of the
  universe.}, and its centre-of-mass position $\vec{x}$.
We do not include neutrinos in the mass calculation. Nonetheless,
we have checked that including neutrino particles in the mass
count does not affect our final result.

In the matching of halos between TianNu and TianZero, we define a
matched pair if the coordinates of the two halos are separated
by less than 100 kpc/$h$ in their respective volumes, and if
their masses vary by less than 10\%. We are able to associate
98\% of all halos between TianNu and TianZero in this way,
corresponding to a total of 27 million pairs of halos in range
$10^{13}\lesssim m_h/M_\odot\lesssim 10^{15}$.

\begin{figure} \centering
  \includegraphics[width=1.0\linewidth]{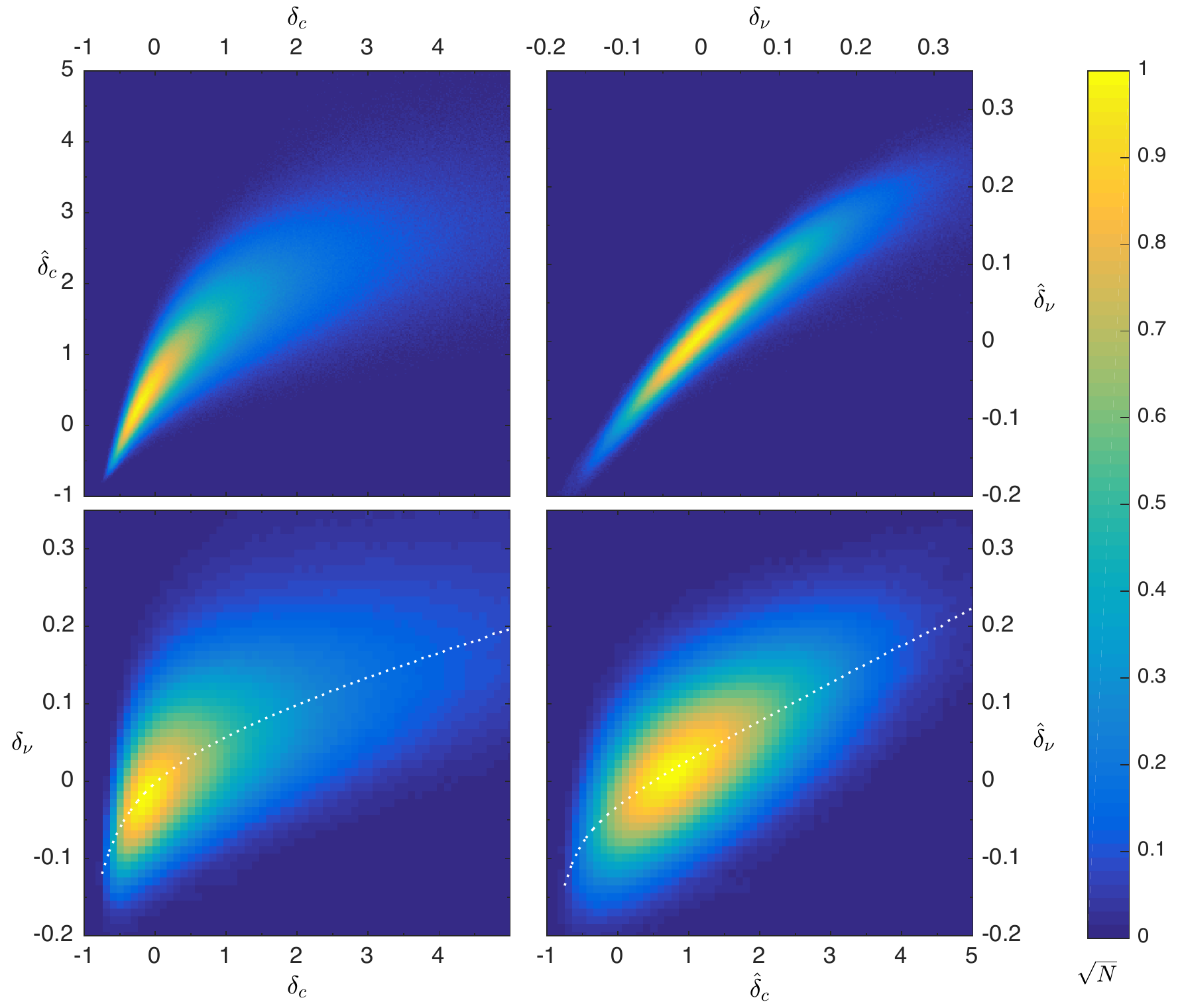}
  \caption{Correlations between simulated and/or reconstructed
    densities for CDM and neutrinos. The top two panels show
    simulation-reconstructed density field correlation
    $\hat\delta_c-\delta_c$ and
    $\hat\delta_\nu-\delta_\nu$. Bottom two panels show
    neutrino-CDM density correlations for simulated fields
    $\delta_\nu-\delta_c$ and reconstructed fields
    $\hat\delta_\nu-\hat\delta_c$.}
  \label{fig.7}
\end{figure}
\paragraph{Density reconstruction and correlation.}
Here we analyze the degree to which $(\delta_c,\delta_\nu)$
can be reconstructed using linear theory. We compute local
values of $\delta_c$ and $\delta_\nu$ within cubes of width
9.4 ${\rm Mpc}/h$ centered on the position of each halo in TianNu.
We could have used the exact values of $\delta_c$ and
$\delta_\nu$ extracted from the neutrino simulation, 
however these quantities are not directly observable.
Instead, we linearly reconstruct both densities using
the TianNu halo density field. This is done in an analogous
manner to the velocity reconstruction in \cite{2015PhRvD..92b3502I}.
Halos are cloud-in-cell (CIC) interpolated to the same
grid we used for particles to obtain the halo density
field $\delta_h$. Then in Fourier space, we compute the
temporary reconstructed fields
\begin{equation*}
  \hat\delta^\dagger_i(k) = \frac{T_i(k)}{T_c(k)} \frac{1}{b(k)} \delta_h(k),
\end{equation*}
where $i=c,\nu$ stands for CDM or neutrinos. $T_i(k)$ is the
transfer function in Fourier space, and the bias
$b(k)=P_{ch}(k)/P_{cc}(k)$, the ratio of CDM-halo cross power
spectrum and CDM auto power spectrum measured in the simulation.
Then we apply a low pass Wiener filter $W(k)$ to each of these
$\hat\delta^\dagger_i(k)$ to get the reconstructed field
$\hat\delta_i(k)=\hat\delta^\dagger_i(k)W(k)$,
where the $W(k)$ is computed as:
\begin{equation*}
  W(k) = \frac{B(k)P_{ss}(k)}{P_{ss}(k)+N^2(k)}
\end{equation*}
where $N^2(k)=P_{ss}(k)-B^2(k)P_{rr}(k)$ and
$B(k)=P_{rs}(k)/P_{rr}(k)$. Here $P_{ss}$ and $P_{rr}$ are the
auto power spectra of the simulated and reconstructed density fields,
respectively, and $P_{rs}$ is their cross power spectrum.
An inverse Fourier
transform gives us the density contrasts. We then proceed
to cubically average these $\hat\delta$'s about the halo
position so the resultant density field is constant over
scales of $9.375$ ${\rm Mpc}/h$ (9 grid cells). We label
the result $\hat\delta_c$ and $\hat\delta_\nu$ and
stress to the reader that these are observable quantities.  We have
checked that computing $\hat\delta_c$ and $\hat\delta_\nu$
from TianZero halo catalog does not affect our result.
The simulated density fields are computed via CIC interpolation of
particles onto a uniform mesh containing $1152^3$ cells and then
filtered in Fourier space by the same Wiener filter $W(k)$. Here we
just label them $\delta_c$ and $\delta_\nu$.

In Fig.6 we show halo number density as functions of their
reconstructed and/or simulated density fields. The top two panels show
the quality of our reconstruction method. Halos are not perfectly
distributed on single diagonal lines, but ideally reflect the
correlation between simulated and reconstructed $\delta_c$'s or
$\delta_\nu$'s.  The Pearson's $r$
for $\delta_c$ is $\rho_{c,c}=0.80$ and for $\delta_\nu$ is
$\rho_{\nu,\nu}=0.95$. It is more difficult to predict $\delta_c$
since halos are modelled as unit point mass in the reconstruction
and any sub-halo structures (1-halo term) are neglected.
This directly leads
to a suppression and large scatter in $\hat\delta_c$ when
$\delta_c\gg 1$. $\delta_\nu$ is less affected by this as neutrinos do
not contain much structure on small scales due to their thermal
velocities. The bottom two panels show $\delta_c-\delta_\nu$
correlations for simulation and reconstructed fields. The right one
is identical to Fig.2 in the main text. When we repeat our analysis
with density contrasts from the simulation $(\delta_c,\delta_\nu)$
instead of halo reconstructed ones $(\hat\delta_c,\hat\delta_\nu)$,
we find the best fit will be
$\partial(\Delta r/r)/\partial\delta_\nu= (0.00283\pm 0.00105)
\log_{10}(r/r_{\rm pivot})-(0.00260\pm 0.00023)$,
and $r_{\rm pivot}=4.56\times 10^5$.

\end{document}